\documentclass[12pt]{iopart}
\usepackage{graphicx}
\begin{document}

\title{Propagation of light in low pressure gas. }

\author{Jacques Moret-Bailly}
\address{Retired professor, Universit\'e de Bourgogne \\ 265 rue St Jean F-21850 St Apollinaire France.}
\ead{jmo@laposte.net}
\begin{abstract}

The criticism by W. E. Lamb, W. Schleich, M. Scully, C. Townes  of a simplified quantum electrodynamics which represents the photon as a true particle is illustrated. Collisions being absent in low-pressure gas, exchanges of energy are radiative and coherent. Thin shells of plasma containing atoms in a model introduced by Str\"omgren are superradiant, seen as circles possibly dotted.
Spectral radiance of novae has magnitude of laser radiance, and column densities are large in nebulae: Superradiance, multiphoton effects, etc., work in astrophysics. The superradiant beams induce multiphotonic scatterings of light emitted by the stars, brightening the limbs of plasma bubbles and darkening the stars. In excited atomic hydrogen, impulsive Raman scatterings shift frequencies of light. Microwaves exchanged with the Pioneer probes are blueshifted, simulating anomalous accelerations. Substituting coherence for wrong calculations in astrophysical papers, improves results, avoids ``new physics''.
\end{abstract}


\section{Introduction}\label{intro}
{\it Many physicists, such as designers of gas lasers, will think this article is unnecessary since many experiments confirm that the theory of optical coherence introduced by Einstein \cite{Einstein1917} is verified in low pressure gas by many observations. But other physicists introduce photons to obtain incoherent interactions of light with atoms without collisions.

W. E. Lamb, W. Schleich, M. Scully, C. Townes \cite{WLamb,WLamb2} put the spectroscopists cautioned against a common error in the use of quantum electrodynamics: the photon results from the quantization of a system of normal modes of the electromagnetic field, modes of which they give examples.

But the photons they call ``pseudo-particles'' are still often seen as small particles independent of choice of an optical system.
The purpose of this paper is to illustrate the work of  Lamb et al. by recalling some essential properties of electromagnetism, then giving some other examples of correct and successful use of coherent spectroscopy.}

\medskip
The introduction of a complex permittivity in Maxwell's equations gives a phenomenological, effective representation of the interaction of light with matter: refraction by the argument of the complex refractive index, absorption or emission by varying its module, resonances explaining the variations of permittivity by coupled oscillators.

The confrontation between Einstein and Planck on the origin of the quantization of energy exchanges between atoms and light ended when Planck accepted the interpretation of the photoelectric effect given by Einstein. This interpretation is questionable now because the interaction of light with the surface of a metal is complex. This interpretation was at the origin of the term photon, specified by quantum electrodynamics (QED).

QED introduces the photon by quantization of the energies of "normal modes" of the electromagnetic field. W. E. Lamb \cite{WLamb}, W. E. Lamb, W. Schleich, M. Scully, C. Townes \cite{WLamb2} emphasize an usual lack of rigor in the use of the photon because the normal modes are often not defined. Consequently, the risk of error  is so large that they recommend to avoid the use of the word ``photon''. They give examples of definitions of these normal modes, rather than their general definition.

To define ``normal modes'', Section \ref{modes} starts in subsection \ref{Maxwell} by a study of the preservation of the linearity of Maxwell's equations in vacuum while matter is introduced.

This linearity is needed to introduce the optical modes in subsection \ref{omodes}. Subsection  \ref{nmodes} defines the normal modes of a well defined optical system.

\medskip
In an article that shows that the irradiance of an atom by a distant star decreases with distance, astrophysicist Menzel \cite{Menzel} wrote:   {\it The so-called ``stimulated'' emissions which have here been neglected should be included where strict accuracy is required. It is easily proved, however, that they are unimportant in the nebulae.} This conclusion seems a consequence of a confusion between radiance and irradiance, quantities which have long been represented by the same symbol $I$. Of course, the irradiance of each atom decreases with distance, but the number of involved atoms increases because the surface of a section of a beam of interacting light increases. Thus the variation d$I$ of radiance of a light beam along a path d$x$ is given by the formula d$I = BI$d$x$ where $B$ is an Einstein coefficient \cite{Einstein1917} which depends only on the crossed medium.

Long physicists had the same view as Menzel, the best of them saying ``Townes' maser will not work''. Now, they use Einstein's theory of optical coherence \footnote{In this paper, ``coherence'' means ``space-coherence'' : identical molecules located on a wavefront of an electromagnetic wave emit identical waves. The radiation of a temporally sine wave is named ``time-coherence''.}  and older laws of physics, the applications of which are commonly verified with lasers: The pressure of plasma in gas lasers must be low enough that incoherent interactions, sources unacceptable losses of energy, are negligible.

 Section \ref{coherence} reminds how, at higher pressures, for instance in the atmosphere, both coherent and incoherent Rayleigh scatterings are observed.

Section \ref{compa} explains the practical use of coherent and incoherent computations

Section \ref{Stromgren} applies coherent spectroscopy to a {\it model} defined by astrophysicist Str\"omgren: an extremely hot source is placed in low pressure hydrogen, cold at great distances. The nature of the source (which may be a supernova) is not important. Hydrogen is assumed pure.

Section \ref{applications}  justifies several theories done in astrophysics, theories rejected by authors whose spectroscopy did not take the coherence into account. It suggests other applications.

\section{Optical modes.}\label{modes} 
\subsection{Linearity of Maxwell's equations.}\label{Maxwell}
Maxwell's equations in vacuum are linear. Matter may be introduced in two ways:

- Usually, matter has a phenomenological representation: the permittivity in the vacuum $\epsilon_0$ is multiplied by a relative permittivity $\epsilon$  which may depend on the field, so that Maxwell's equations are often not anymore linear. A nonlinear permeability may be introduced too.

- We can compute the electromagnetic ``retarded field'' emitted by an electric charge. Changing the sign of time, Maxwell's equations are invariant, the retarded field becomes an ``advanced field'' absorbed by the charge. Schwarzschild and Fokker change also the sign of the charge, obtaining the ``conjugate field''. Adding a retarded field and its conjugate field, the charges cancel, so that the conjugate field generates the retarded field: Thus, in the initial problem, the introduction of the charge may be replaced by the introduction of the conjugate field. Maxwell's equations remain linear, the introduction of the charge is replaced by convenient conditions at the limits, in the past.

The norm (or intensity) of a solution of linear Maxwell's equations is its energy computed with the hypothesis that there is no other field in the universe. The scalar product of two solutions is deduced from the norms of each and of their sum. Any solution may be developed on a complete set of orthogonal solutions.
\footnote{A common mistake is an extension of linearity to the field of electromagnetic energy. For instance the energy radiated by an accelerated charge is computed without explicit consideration of external fields. Thus we read that the electron of the classical hydrogen atom radiates energy and falls to the nucleus. In the stationary states, by correcting the classical elliptic orbit slightly to reflect the ``Lambshift", the interference of the radiated field and the zero point field produces a zero radiation of energy.}

\subsection{Modes of a linear field.}\label{omodes}
The linearity of Maxwell's equations means that all linear combinations of solutions with real multiplicative constants, is a solution. Thus the solutions are represented by the points of a vector space $\bf \Sigma$.

A strict mode is a set of all solutions which differ by a multiplicative real constant. It is represented by a ray of space  $\bf \Sigma$.
Scalar product, orthogonality are transferred from the solutions to the modes.

\medskip
The word ``mode`` is often ambiguous. Prefixes may specify it:

- s-mode for the strict modes.
 
- g-mode, that is ``geometrical mode'' for the set of all s-modes which differ by an addition of a constant to the time $t$.

- us- or ug-mode for a set of two orthogonal s- or g-modes which differ only by the polarization.

A continuous monochromatic wave propagating in a diffraction-limited light beam, is in a s-mode.

A light pulse  whose spectrum corresponds to its Fourier transform, propagating in a diffraction-limited beam, defines a s-mode. The set of these s-modes, makes a g-mode.

Photometry uses sets of us-modes modes making ug-modes.

\subsection{Normal modes.}\label{nmodes}
The concept of normal mode comes from acoustics. When a musical instrument is used "normally", the frequency spectrum emitted does not depend on the intensity of the emission, the emissions occur in well-defined s-modes.
This property results from a linear approximation of the operation of the instrument and a replacement of non-linearities by conditions at the limits: for example, the non-linearities which create and transfer the energy of sound at the ends of a flute are replaced by the hypothesis of a constant pressure. For a too large excitation of the flute, the non-linearities at its ends cannot anymore be represented by conditions at the limits. If it remains possible to define modes, these modes are not anymore ``normal'': the set of normal modes is replaced by an other set of orthogonal modes.

It is possible to find an equivalent of acoustic normal modes in optics: 

i) In large, homogeneous regions, orthogonal modes are defined using Schwarzschild-Fokker representation of the charges. 

ii) These modes are physically stabilized by conditions at the limits introduced by phenomenological representation of regions. For instance: the tangential component of the electric field is null on a mirror.

Willis E. Lamb \cite{WLamb} shows such examples of optical normal modes.

{\it The definition of normal modes involves: i) a choice of a set of orthogonal modes; ii) a physical system which forbids a mixture of any of these modes with another mode.}

\section{Coherence and incoherence.}\label{coherence} 
\subsection{Huygens' construction.}
The theory of light propagation in the atmosphere illustrates coherent and incoherent scatterings well:

Huygens' construction shows the propagation of monochromatic waves in a homogeneous continuous medium, considering that each infinitesimal fragment of the medium located on a surface wave emits a monochromatic wavelet of same frequency, coherent with the exciting wave.

 Einstein \cite{Einstein1917} assumes that identical molecules constituting an homogeneous medium have a determinist behavior, so that they emit wavelets whose amplitudes are the product of the incident amplitude by a same complex number \footnote{Heavy molecules may be fluorescent, with an unpredictable time constant; this is an uncommon exception.}. Thus Huygens' construction applies to particles of a real medium if the finite density of particles in the vicinity of a wave surface is large enough: a coherent wave is scattered. In a transparent medium, the scattered wave is usually delayed by $\pi/2$. The identity of the initial and scattered wavefronts allows an interference of the two waves to produce a single monochromatic, refracted wave.

The validity of the old theory of refraction is extended to the emission or absorption of light in a real, discontinuous medium: by a convenient scattering, the complex amplitude of an incident wave is multiplied by a complex ``amplification'' coefficient, preserving the wave surfaces and respecting the laws of thermodynamics. This extension is not obvious because the energy exchanges with the molecules are quantized, so we must admit the existence of a process of (de-)coherence, either quantum or classical. This process uses the fluctuations of the absolute field to select the atoms which perform a transition.

\medskip
Introduction of matter by Huygens' theory can be faulty if the scattering of light depends on a stochastic parameter.
It is the case of colliding molecules which are fast evolving, unpredictable systems. In the neighborhood of a critical point of a gas, all molecules are in these unpredictable states, so that Huygens' theory is no longer valid. Away from the critical point, most molecules are free, the coherent Rayleigh scattering (giving refraction) and incoherent Rayleigh scattering (blue of the sky ...) become compatible.

In low pressure gas almost all collisions are binary, the density of which is proportional to the square of the pressure. To avoid incoherent interactions, the pressure in a gas laser must be of the order of 100 Pascals. As the pressure in the nebulae is generally much lower, all light-matter interactions are coherent in nebulae.
This conclusion is exactly opposite to Menzel's statement: 

{\it All interactions of light with low pressure gases are coherent.}

\section{Using coherent or incoherent propagation of light}\label{compa}

\subsection{Using Einstein's theory.}\label{Einstein}
Clausius' optical invariant (or optical extent) $C$, in a medium of index of refraction $n$, is defined for beams consisting of light rays depending continuously on two infinitesimal parameters. A section of the beam by a surface $S$ is represented by a vector $\rm d\bf S$ normal to $S$ and whose length measures the crossed surface. Set d$\Omega$ a vector colinear to a ray, whose length measures the solid angle which contains the directions of the rays, such that $C=n^2{\rm d}{\bf S}.{\rm d}{\Omega}$ is positive. In a transparent medium, $C$ does not depend on the section. The flux of energy in the beam is written $IC$ where $I$ is the radiance of the beam (us-mode). The definition of $I$ applies to the whole spectrum (energetic radiance), or to an infinitesimal set of frequencies  (spectral radiance). Here, we use only spectral radiances. $I$ is constant in a transparent medium.

Einstein showed that the variation of $I$ along a path d$x$ is:
\begin{equation}
  {\rm d}I=BI{\rm d}x\label{EqEi}
\end{equation}
where $B$, named Einstein's $B$ coefficient depends only on the state of the gas. Here, $B$ is assumed isotropic.
If $B$ is negative, the equation is the usual equation of absorption; else it represents a coherent amplification.

The integration of formula \ref{EqEi} introduces a constant for which two values are generally used:

- $I$ may be set relative, equal to zero in a blackbody. The neglected zero point field (Planck \cite{Planck1911},Einstein and Stern \cite{Einstein1913}) is introduced through Einstein $A$ coefficient.

- $I$ is set to its absolute value $h\nu^3/c^2$ in a blackbody, that is, for light at usual temperatures, in the dark. Einstein coefficient $A$ equals zero.

An advantage of the use of absolute radiance and field is the correctness of the calculation of the energy exchanged with matter, which involves calculating the variation of the square of the absolute field which differs from the variation of the square of a relative field. The error is frequent in ``photon counting'' experiments.

Diffraction (and polarization) limited g-modes of interest in astrophysics propagate in beams limited by the aperture of a telescope and the diffracted image of a distant point. The optical extent of these beams is the square of the wavelength $\lambda^2=c^2/\nu^2$. In these beams, an infinite, polarized sine wave defines an s-mode; each pulse of polarized natural light whose frequency spectrum $\Delta\nu$ corresponds to the pulse duration, defines also a s-mode. Thus, the absolute energy in a s-mode is obtained multiplying the radiance $I_\nu$ by $\lambda^2/2=c^2/2\nu^2$. Corresponding to a degree of freedom, it is equivalent to $kT/2$ at high temperature, as required by thermodynamics. 

\medskip
Consider a transition between two nondegenerate levels of potential energy $E_1$ and $E_2>E_1$, whose respective populations are $N_1$  and $N_2$. Boltzman's law defines a transition temperature $T_B$ such that $N_2/N_1 =exp[(E_1-E_2)/kT_B]$. By convention, negative values of $T_B$ are accepted. In homogeneous enough environments, it is agreed to treat separately refraction and change of radiance. With this convention, a ray is attenuated or amplified by the medium which it crosses without its geometry is changed. Only the initial phase of a ray of unknown origin must be considered as stochastic.

If the medium is opaque, $T_P$ equals $T_B$. Otherwise, $T_P$ tends to $T_B$. The algebraic value of the coefficient of amplification may be computed from Einstein's coefficient $B$, the sign of which is positive if $T_B<0$ or if $T_B>T_P$.

The computations of refraction and amplification/absorption use the same equation: the images are preserved by amplification/absorption.

\subsection{Incoherent interactions.}
The results of the interaction of a neutron with an atom of uranium are numerous, only statistically predictable. Thus, the result of each interaction is drawn. It is useless to introduce the pilot wave of the neutrons because the wavelength is so small that the trajectory of the neutrons is straight between the collisions, and the possibility of interference is negligible. With a large number of interactions, the result is good.
\medskip

The interaction of light with a water drop of a cloud depends on the size of the drop, it involves refraction and diffraction. With variable distances between drops, the interference of light beams may generally be ignored: Thus Monte-Carlo computations explain generally well the propagation of light in the clouds, more generally in opalescent media. However they do not work to explain particular observations involving interferences, such as rainbows.

\medskip
Quantum mechanics associates a particle with a linear ``Schr\"odinger wave'' $\Psi$, scalar field of complex value whose Hermitian square at a point is proportional to the probability of finding the particle at that point. It is not known how to define a $\Psi$ wave for the photon, so a scalar function of the electromagnetic field is used in its place. Associating photons to this wave requires the definition of ``normal modes" valid only in a limited optical system out of which the photons can not be exported. The pseudo-particle photon must be used with a great care \cite{WLamb,WLamb2}  that many physicists do not have. The astrophysicists apply Monte Carlo calculations to photons, to study light propagation in nebulae \cite{Zheng, Nilsson}, while, without collisions, the stochastic behavior producing the blue of the sky and pointed by the word ``Monte-Carlo'' does not exist.

\medskip
Summarize the study of the propagation of light in a low pressure gas, in spectroscopy and in astrophysics:

In standard spectroscopy using coherence, light propagates in light beams whose flux of energy is measured by the radiance. In a transparent medium, radiance is constant, even far from a waist of the beam where the density of energy may be very low. The variation of the radiance results from Einstein's $B$ coefficient depending from the state of the gas. {\it During a nanosecond pulse of ordinary incoherent light, the collisions are negligible so that all exchanges of energy are radiative and coherent.}

In usual astrophysics, light is represented by photons packets. These photons are not defined by normal modes. They propagate along a light ray, but they do not have a pilot wave, they cannot interfere. The light-gas interactions are studied numerically using many types of codes mixing generally propagation of light and evolution of the gas \cite{Iliev1}. In some studies, the flux of photons is similar to a radiance \cite{Abel}, in others a photon-conserving rule is used. Iliev et al. observe that ``the agrement between the various codes is satisfactory although not parfect''. This is very surprising for a spectroscopist.

\medskip
Ignoring the optical coherence, astrophysicists are depriving themselves of a simple and powerful tool for interpreting observations of nebulae.

\section{Study of an astrophysical size model introduced by Str\"omgren.}\label{Stromgren}
\subsection{Str\"omgren's main results.}\label{results}
\begin{figure}
\includegraphics[width=15cm]{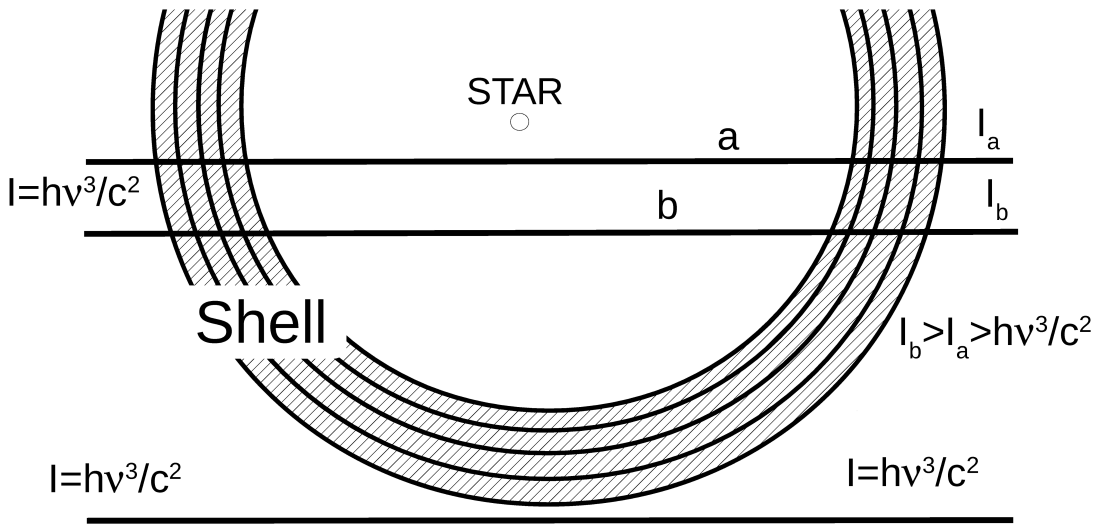}
\caption{Comparison of the amplifications of rays crossing a Str\"omgren shell: the path inside each infinitesimal shell is larger for ray b than for ray a, null for ray c. Thus, versus the distance of the ray from the star, the total amplification increases, then falls, therefore has at least a maximum.}
\label{coq}
\end{figure}

Str\"omgren \cite{Stromgren} defined a model consisting of an extremely hot star immersed in a vast, low-density and initially cold hydrogen cloud. The ultra-violet emitted by the star ionizes almost completely the hydrogen of a {\it Str\"omgren's sphere} which becomes transparent. Traces of atoms appearing in the outer regions of the sphere absorb energy by collisions and from light emitted by the star at their own eigen-frequencies. They dissipate this energy by radiating spontaneously into all directions.

\begin{figure*}
\includegraphics[width=4cm]{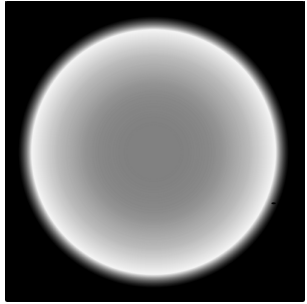}
\caption{Appearance of a Str\"omgren's shell, supposing that spontaneously emitted light is neither absorbed nor amplified.}
\label{sphere}
\end{figure*}
 This energy dissipation lowers the temperature and causes a catastrophic increase in the density of atoms, so that the sphere is surrounded by a {\it Str\"omgren's shell} relatively thin, radiating intensely.

\begin{figure*}
\includegraphics[width=4cm]{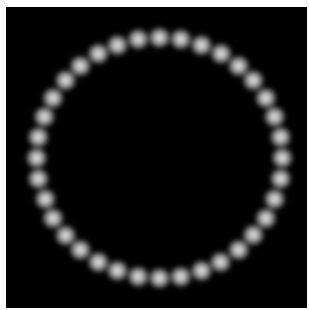}

\caption{With a strong superradiance a Str\"omgren's shell appears as a pearls necklace.}
\label{mode}
\end{figure*}

\subsection{Superradiance.}\label{superradiance}
 Figure \ref{coq} shows the amplification of a light ray by the Str\"omgren's shell which may be split into infinitesimal, concentric, spherical shells: For rays having crossed near the star, the paths inside all infinitesimal crossed shells vary little. Farther, these paths grow faster and faster up to a maximum amplification. Finally, the number of crossed infinitesimal shells decreases, the amplification falls to reach no amplification. Figure \ref{sphere} shows the appearance of the system supposing that the spontaneously emitted light is nether absorbed nor amplified.

Str\"omgren did not know the intense, coherent interactions of light with matter. It is generally assumed that the atomic hydrogen of the nebulae absorbs light ``on the spot'', that is is opaque at the resonance frequencies.
The plasma in the shell is similar to the plasma in gas lasers: the amplification factor is high, resulting in intense superradiance, maximal for rays located at a distance $R$ from the center of the sphere. $R$ defines precisely the inner radius of the shell. By competition of modes, these rays emit most of the available energy. Into a given direction they generate a circular cylinder.

If the superradiance is large, the competition of modes playing on these rays selects a system of an even number of orthogonal modes, as in a laser whose central modes have been extinguished, for example by an opaque disc (daisy modes). Thus, strongly superradiant rays of a transition form a circle regularly punctuated (Fig 3), and the atoms are quickly de-excited.
With a lower superradiance, or by a mixture of several frequencies in a wide band detector, the ring may appear continuous.

\subsection{Multiphotonic scattering}\label{multiphoton}
\medskip
The spectral radiance of the rays emitted by a supernova has, at each frequency, the radiance of a laser, so that multiphoton interactions and transitions between virtual levels are allowed. All frequencies may be involved in simple combinations of frequencies which result in resonance frequencies of hydrogen atoms. Thus, an important fraction of the energy emitted by the star is absorbed. All multiphoton absorptions and induced emissions form a parametric induced scattering that transfers the bulk of the energy of the radial rays emitted by the star to the superradiant rays. If the star is seen under a solid angle much smaller than the points forming the ring, it is no longer visible.

\subsection{Frequency shifts.}\label{shifts}
A large coherent transfer of energy between two co-linear light beams is difficult because the wavelengths being generally different, the difference of phase between the two beams changes, so that the transferred amplitudes cancel along the path. To preserve the coherence, the usual solutions are: the use of two indices of refraction of a crystal, the use of non-colinear broad light beams, or the use of light pulses:

In the ``Impulsive Stimulated Raman Scattering'' (ISRS), the use of short laser pulses limits the phase shift of two light beams of slightly different frequencies, so that their coherence may be preserved: G. L. Lamb \cite{GLamb} wrote that the length of the used light pulses must be ``shorter than all relevant time constants''. A time constant is the collisional time because collisions break the phases. An other is  Raman period which produces the period of beats between the exciting and the scattered beams.
 
\medskip
In appendix A, a general description of ISRS is avoided because the studied ``Coherent Raman Effect on Incoherent Light'' (CREIL) may be simply obtained by addition of a Raman coherent scattering to the Rayleigh coherent scattering which produces the refraction. For a ``parametric effect'', that is to avoid an excitation of matter so that matter plays the role of a catalyst, several light beams must be involved simultaneously. With the exception of frequency shifts, the properties of the CREIL are those of refraction:

- The interactions are coherent, that is the wave surfaces are preserved.

- The interaction is linear versus the amplitude of each light beam so that there is no threshold of amplitude.

- The entropy of the set of light beams increases by the exchange of energy which shifts their frequencies.

- Locally, the frequency shift of a beam is proportional to the column density of active scattering molecules and it depends on the temperatures and irradiances at all involved light frequencies.

- Lamb's conditions must be met. 

\medskip
Using ordinary incoherent light made of around one nanosecond pulses, the pressure of the gas and a Raman resonance frequency must be low, so that the effect is weak. With atomic hydrogen, frequency 1420 MHz of the hyperfine transition in 1S state is too large; in the first excited state the frequencies 178 MHz in the 2S$_{1/2}$ state, 59 MHz in 2P$_{1/2}$ state, and 24 MHz in 2P$_{3/2}$ are as large as allowed, very convenient.

\medskip
The background radiation is always implied. As it is generally nearly isotropic, a lot of beams is implied, their irradiance is large. Thus, for light, there is always a redshift component.
 
These frequency shifts are easily observed in laboratories using femtosecond laser pulses, or, with longer pulses, in optical fibers. The frequency shift in a gas is roughly inversely proportional to cube of the length of the pulses, so that astronomical paths are needed with usual light.

\begin{figure*}
\includegraphics[width=8cm]{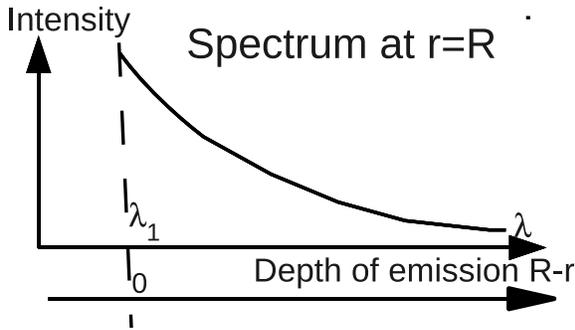}
\caption{Theoretical, qualitative spectrum of light spontaneously emitted along a ray crossing a Str\"omgren sphere. Observed at the surface of the sphere (r=R), the maximum of radiance is at the laboratory wavelength ($\lambda_1=\lambda_0$). In the Str\"omgren's shell, all wavelengths decrease, the scale of the spectrum is changed: $\lambda_1<\lambda_0$. The spectrum depends on the distance $\rho$ between the ray and the star.}
\label{caj}
\end{figure*}

\medskip
The density of atomic hydrogen is negligible in a Str\"omgren's sphere except near the surface, where it grows with an exponential look to the surface. Thus, the intensity of ``spontaneous emission'' is low in depth, high in surface. In propagating in a medium that contains excited hydrogen atoms, a light beam provides energy to thermal radiation of high irradiance, and receives energy from the hot rays emitted by the star whose irradiance is low. It is reasonable assuming that the balance is negative, so that, at the surface, the weak, deep emission is redshifted, while the stronger, surface emission is at the laboratory wavelength $\lambda_0$  (Fig. \ref{caj} with $\lambda_1=\lambda_0 $, laboratory wavelength).

\medskip

In the shell, near the sphere, it remains excited hydrogen able to catalyze exchanges of energy. The energy emitted by the star which propagates radially at speed $c$ is transferred to the tangent rays whose {\it radial component of speed} is low, so that the irradiance by warm rays becomes very large: the cold spontaneous emission receives energy and the spectrum is shifted towards shorter wavelengths  (Fig. \ref{caj} with $\lambda_1<\lambda_0$).

\section{Possible applications in astrophysics.}\label{applications}
\subsection{The distorted Str\"omgren's sphere of supernova remnant SNR1987A}

\begin{figure*}
\includegraphics[width=15cm]{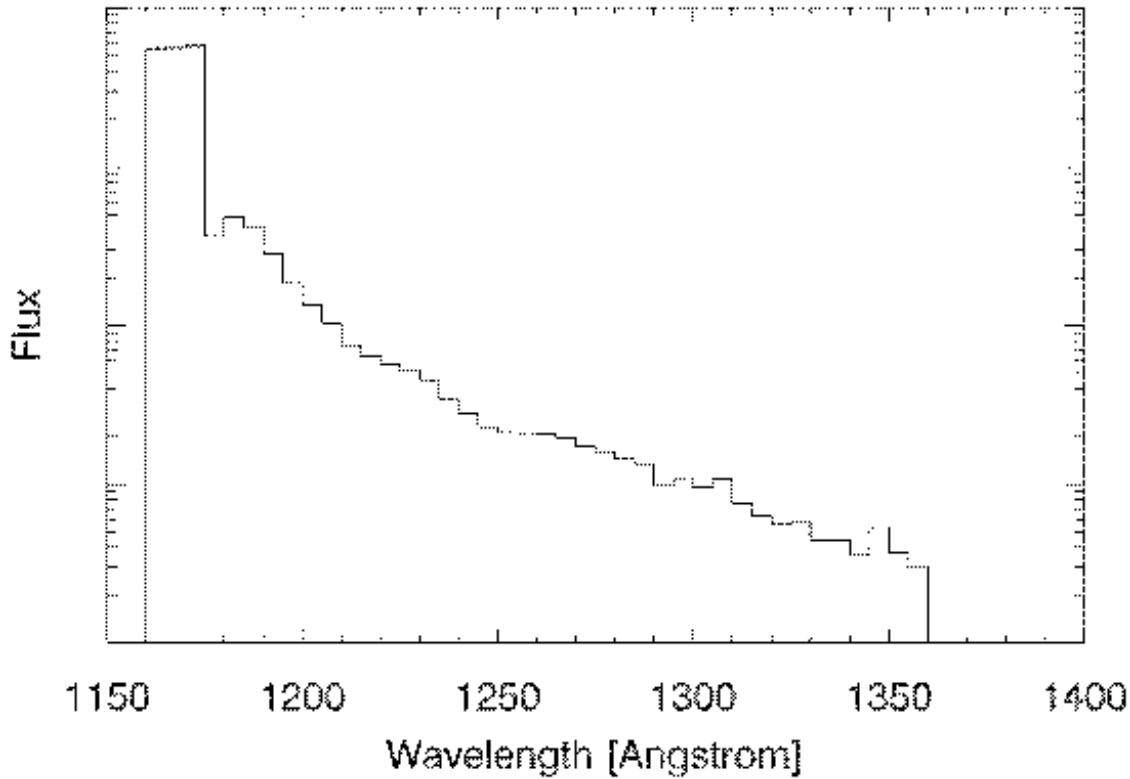}
\caption{Spectrum of the spontaneous emission of the disk inside the main necklace of SNR1987A, computed by Michaelet al.\cite{Michael}.}
\label{camic}
\end{figure*}

\begin{figure*}
\includegraphics[width=15cm]{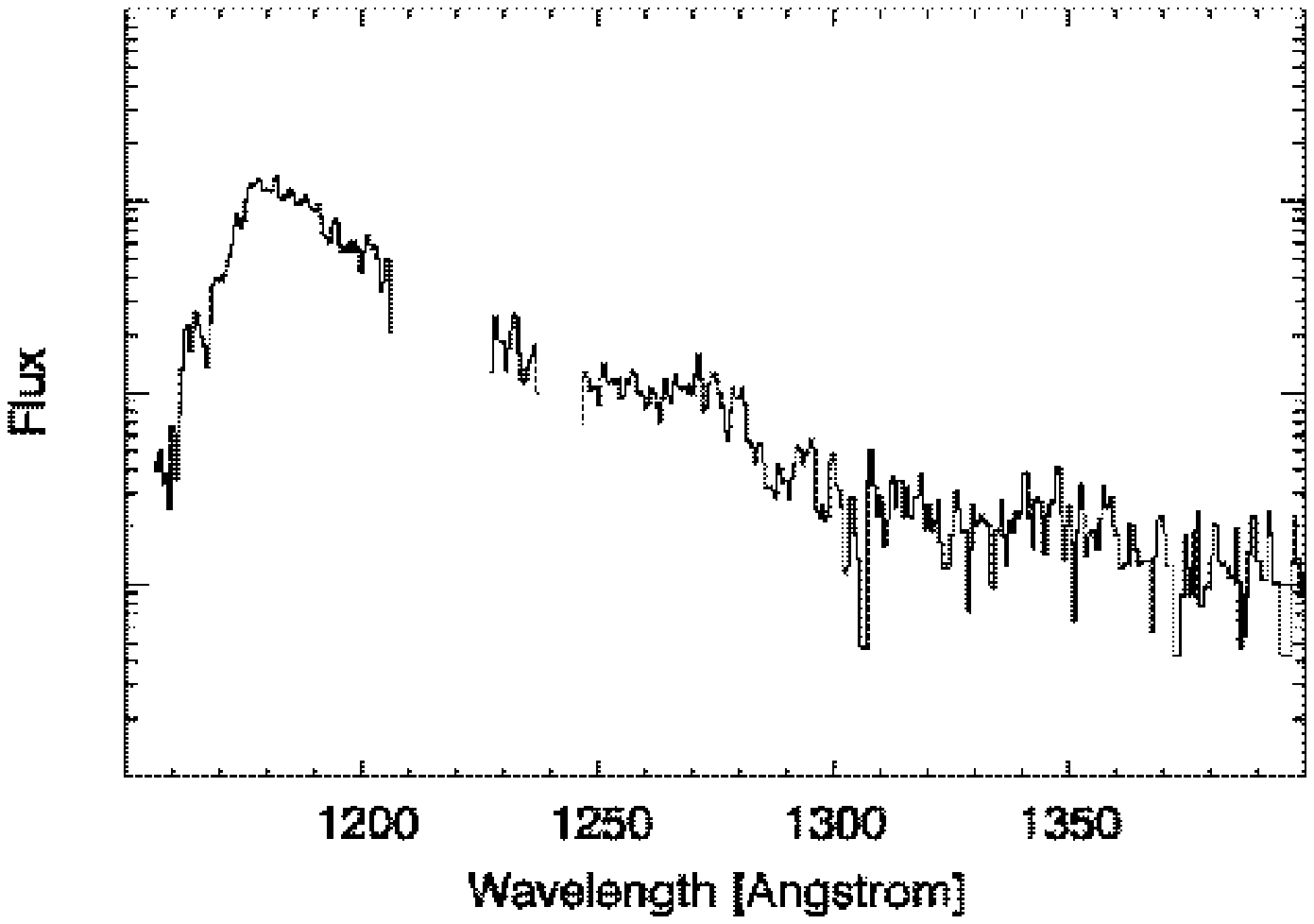}
\caption{Spectrum recorded inside the main ring of SNR1987A \cite{Michael}.} 
\label{recs}
\end{figure*}

The  supernova remnant SNR1987A is surrounded by relatively dense clouds of hydrogen making a ``hourglass'' \cite{Sugerman}. These clouds have been detected shortly after the explosion by photon echoes. Burrows et al. \cite{Burrows} criticized the interpretation of the three rings of SNR1987A as the limbs of the hourglass because, without superradiance, these limbs are wide as in figure \ref{sphere}. Assuming that the hourglass is a strangulated Str\"omgren's sphere, and taking the superradiance into account, distorted figures similar to figure \ref{mode}, represent the three ``pearl necklaces'' of SNR1987A. Evidently it is necessary to take into account several lines of hydrogen, variations of the density of hydrogen, other lines excited by hydrogen, so that the monochrome image of SNR1987A is complicated.

\medskip
Our spectrum of light emitted inside the Str\"omgren sphere (Fig. \ref{caj}) does not show a strong peak as the calculated (cut by the recorder) spectrum by Michael et al.\cite{Michael} (Fig. \ref{camic}). Our computation is better than Monte-Carlo, but the starting point is the same: the shifts of the Lyman $\alpha$ line are assigned to the propagation of light in the plasma of hydrogen rather than an expansion of the universe, a Doppler effect of winds,... . Probably Michael et al. did not insist on this point because of hostility against any discussion on the origin of the ``cosmological'' redshifts.

The spectrum (Fig. \ref{recs}) emitted within the central ring of SNR1987A results from the superposition of spectra observed at different distances $\rho$ from the center of the disk, so that different spectra corresponding to different paths are added.

\subsection{Overview of other possible applications.}
Many ``planetary nebulae'' show arcs of circles or ellipses punctuated or not: For instance, the ``necklace nebula\textquotedblright{} (PN G054.2-03.4) shows a ring  similar to SNR1987A inner ring, the ``Bubble nebula\textquotedblright{} (SNR 0509) shows a sphere having a continuous bright limb; they may be images of Str\"omgren's systems. The usual explanation of simpler systems of dots (Einstein cross PGC 69457) is that the image of a very bright, distant star is distorted, multiplied by the gravitational lensing of an interposed, dark, heavy star. This explanation has been criticized because it involves a large number of alignment of proper stars with Earth, and because it is difficult to justify a certain regularity of punctuation, the always even number of dots. By superradiance, the phases of two contiguous dots are opposite, so that this feature can be tested by interference if the necklace is incomplete. The spectra could distinguish the two types of rings: the spectrum of the limb of a Str\"omgren's sphere is a line spectrum while the spectrum of a very far, bright star is probably a continuous emission spectrum.

Observed lines of many atoms have the shape of figure \ref{caj}. They may be generated by atoms heated in a cloud of hydrogen plasma.

\medskip
The frequency shifts by CREIL effect have many other applications \cite{MBIE,MB0507141,MBAIP06}:

 - Increase in the frequency of the radio-waves exchanged with the Pioneer 10 and 11 probes, resulting from a transfer of energy from the solar radiation where the solar wind is cooled enough to generate atoms. This frequency shift is usually interpreted by Doppler effect, as an anomalous acceleration of the probes.
 It would be interesting (but difficult) to verify that the frequency of a signal is not increased while the frequency of the carrier is.

 - The frequency shifts of the extreme UV lines emitted by the Sun and observed by SOHO are usually attributed to a Doppler effect due to vertical speeds of the sources. But the frequencies observed at the limb are not the laboratory frequencies. Assume that at high pressure and temperature, hydrogen is in a state similar to a crystal, so that a CREIL effect is possible. The paths, thus the frequency shifts, from the depth that emits a line are larger at the limb than at the center so that the laboratory frequencies are preserved.

 - Neutron stars are never observed inside a cloud of hydrogen. The theoretical spectrum of such ''accreting neutron stars'' heated to a very high temperature by accretion of a cloud of hydrogen is very similar to the spectrum of a quasar, including the Karlsson's periodicities. Thus the quasars may be in our galaxy or close to it, so that they are not enormous and do not move very fast.

 - High redshifts appear where hydrogen is  atomic and excited.

 - It is necessary to re-examine the scales of distances deduced from Hubble's law which assumes a redshift of the spectra proportional to the path of light whereas the example of SNR1987A shows that it is necessary to take account of other parameters: density and state of hydrogen, temperature of the studied rays, temperatures and radiances of the other rays. Thus, an important work appears necessary to press the sponge of the maps of the galaxies. To get some reliable distances, one can use the dynamics of the galaxies to evaluate, without black matter, their sizes thus their distances.

\section{Conclusion}
Obvious but common errors in electromagnetism are recalled: the field of density of electromagnetic energy is not linear. It  is proportional to the square of the {\it absolute} electromagnetic field. The photon is not a real, small particle, it needs to be defined by normal modes. The Monte Carlo calculations must be reserved for the propagation of light in opalescent media.

\medskip
The powerful tools developed in connection with lasers should be used to study the diluted gases present in interstellar space with column densities much higher than those typically obtained in gas lasers.

Superradiance, multiphoton interactions, induced emissions explain:

 - the formation of the "pearl necklaces'' of supernova 1987A;

 - the disappearance of the star that supplies energy when these necklaces appeared. 

The same explanations apply to many circles or arcs, punctuated or not, usually attributed to gravitational lensing in improbable alignments of extraordinary stars.

\medskip
In excited atomic hydrogen, a parametric effect resulting from a combination of spatially Coherent Raman Effects acting on temporally Incoherent Light (CREIL), shifts the frequencies of involved (exciting) light beams, in accordance with thermodynamics. Unexpected results appear:

- Hubble's law is explained by a CREIL exchanging energy between light and the microwave background. Hubble's law does not provide a reliable distance scale as the CREIL depends, in particular, on the density of excited atomic hydrogen that works as a catalyst. The voids of the maps of galaxies seem correspond to small regions containing hot hydrogen.

- The shape of many spectral lines broken at the shortest wavelength is due to a spontaneous emission and redshift of the lines in a hydrogen plasma.

- The ``anomalous acceleration" of Pioneer 10 and 11 results from the attribution to a Doppler effect, of the blueshift of the carrier of microwave signals. This shift is due to an exchange of energy between the solar light and the microwaves, where, between 15 and 20 AU, the solar wind cools into excited atomic hydrogen.

\medskip
Can we accept that usual, ordinary, verified laws of physics replace the marvellous system of new laws of physics built on the hypothesis of the big bang?

 \appendix
\section{Appendix: Frequency shifts of time-incoherent light beams by coherent transfers of energy (CREIL).}
In a gas, the CREIL is a parametric effect resulting from the assembly of several Impulsive Stimulated Raman Scatterings (ISRS), and applied to temporally incoherent radiation. Here, it is explained adding Raman impulsive, coherent scatterings to the Rayleigh coherent scatterings which produce the refraction.

To explain the wave propagation, Huygens deduced from a wave surface known at time $t$, a slightly later wave surface, at time $t+\Delta t$. For that, he supposes that each element of volume contained between wave surfaces relating to times $t-\mathrm dt$ and $t$, emits at the local speed of the waves $c$, a spherical wavelet of radius $c\Delta t$. The set of these wavelets has, as envelopes, the sought wave surface and a retrograde wave surface canceled by the retrograde waves emitted at other times.

Let us suppose that each element of volume considered by Huygens contains molecules which emit also a wavelet of much lower amplitude at same frequency, whose phase is delayed by $\pi/2$ (Rayleigh coherent emission). Both types of emission generate the same wave surfaces, so that their emitted fields may{} be simply added. Are $E_0 \sin (\Omega T)$ the\label{shell} incident field, $E_0K\epsilon \cos(\Omega T)$ the field diffused in a layer of infinitesimal thickness $\epsilon=c\mathrm dt$ on a wave surface, and $K$ a coefficient of diffusion. The total field is:

\begin{equation}
E=E_0[\sin(\Omega t)+K\epsilon \cos(\Omega t)]\label{refr}
\end{equation}

\begin{equation}
\approx E_0[\sin(\Omega t)\cos(K\epsilon)+\sin(K\epsilon )\cos(\Omega t)]=E_0\sin(\Omega t -K\epsilon).
\end{equation}
 
 This result defines the index of refraction $n$ by the identification 
\begin{equation}
K=2\pi n/\lambda=\Omega n/c.\label{index}
\end{equation}

Try to replace in this theory of refraction the Rayleigh scattering by a Raman scattering, with a shifting frequency $\omega$, but no initial phase shift.

Setting $K'>0$ the anti-Stokes diffusion coefficient, equation \ref{refr} becomes:
\begin{equation}  
E=E_0[(1-K'\epsilon)\sin(\Omega t)+K'\epsilon \sin((\Omega+\omega)t)].
\end{equation}

In this equation, incident amplitude is reduced to obtain the balance of energy for $\omega=0$.

\begin{eqnarray}
E=E_0\{(1-K'\epsilon)\sin(\Omega t)+\nonumber\\
+K'\epsilon[\sin(\Omega t)\cos(\omega t)+\sin(\omega t)\cos(\Omega t)]\}.
\end{eqnarray}

$K'\epsilon$ is infinitesimal; suppose that between the beginning of a pulse at $t=0$ and its end, $\omega t$ is small; the second term cancels with the third, and the last one transforms:

\begin{eqnarray}
E\approx E_0[\sin\Omega t+\sin(K'\epsilon\omega t)\cos(\Omega t)]\nonumber\\
E\approx E_0[\sin(\Omega t)\cos(K'\epsilon\omega t)+ \sin(K'\epsilon\omega t)\cos(\Omega t)\\
E\approx E_0\sin[(\Omega+K'\epsilon\omega)t].\label{eq4}
\end{eqnarray}

Hypothesis $\omega t$ small requires that Raman period $2\pi/\omega$ is large in comparison with the duration of the light pulses; to avoid large perturbations by collisions, the collisional time must be larger than this duration.  This is a particular case of the condition of space coherence and constructive interference written by G. L. Lamb \cite{GLamb}.

Stokes contribution, obtained replacing $K'$ by a negative $K''$, must be added. Assuming that the gas is at temperature $T$, $K'+K''$ is proportional to the difference of populations in Raman levels, that is to $\exp[-h\omega/(2\pi kT)]-1 \propto \omega/T$.

$K'$ and $K''$ obey a relation similar to relation \ref{index}, where Raman polarisability which replaces the index of refraction is also proportional to the pressure of the gas $P$ and does not depend much on the frequency if the atoms are far from resonances; thus, $K'$ and $K''$ are proportional to $P\Omega$, and $(K'+K'')$ to $P\Omega \omega/T$. Therefore, for a given gas, the frequency shift is:
\begin{equation}
\Delta\Omega=(K'+K'')\epsilon\omega\propto P\epsilon\Omega\omega^2/T.\label{delom}
\end{equation}
The relative frequency shift $\Delta\Omega/\Omega$ of this space-Coherent Raman Effect on time-Incoherent Light (CREIL) is nearly independent on $\Omega$ and proportional to the integral of $Pc{\rm d}t$, that is to the column density of active gas along the path.

The path needed for a given (observable) redshift is inversely proportional to $P\omega^2$. At a given temperature, assuming that the polarisability does not depend on the frequency, and that $P$ and $\omega$ may be chosen as large as allowed by Lamb's condition, this path is inversely proportional to the cube of the length of the pulses: an observation easy in a laboratory with femtosecond pulses requires astronomical paths with ordinary incoherent light.

\end{document}